\documentclass[aps,prl,twocolumn]{revtex4-1}
\usepackage{mathtools}
\usepackage{xcolor}
\usepackage{graphicx}
\usepackage{amsmath}
\usepackage{amssymb}
\usepackage[hidelinks]{hyperref}
\usepackage{fancyhdr}

\fancypagestyle{firstpage}{ 
	
	\fancyhead{\fontsize{8}{13}\selectfont PHYSICAL REVIEW LETTERS {\bf 128}, 054101 (2022)}
	\fancyhead[L]{}
	\fancyhead[R]{}
	\fancyfoot[C]{\thepage}
}
\fancypagestyle{plain}{ 

\fancyfoot[C]{\thepage}}
\newcommand{\ii}{\mathrm{i}}

\begin{document}

\title{\mbox{Hierarchy of Exact Low-Dimensional Reductions for Populations of Coupled Oscillators}}

\author{Rok Cestnik}
\affiliation{\fontsize{8}{10}\selectfont Department of Physics and Astronomy, University of Potsdam, Karl-Liebknecht-Strasse 24/25, 14476, Potsdam-Golm, Germany}
\email{rokcestn@uni-potsdam.de}
\author{Arkady Pikovsky}
\affiliation{\fontsize{8}{10}\selectfont Department of Physics and Astronomy, University of Potsdam, Karl-Liebknecht-Strasse 24/25, 14476, Potsdam-Golm, Germany}


\begin{abstract}
We consider an ensemble of phase oscillators in the thermodynamic limit, where it is described 
by a kinetic equation for the phase distribution density. We propose an {\it Ansatz} for the circular moments
of the distribution  (Kuramoto-Daido order parameters) that allows for an exact truncation at an arbitrary number
of modes. In the simplest case of one mode, the {\it Ansatz} coincides with that of Ott and Antonsen [Chaos {\bf 18}, 037113 (2008)]. 
Dynamics on the extended manifolds facilitate higher-dimensional behavior such as chaos, which we demonstrate with a simulation of a Josephson junction array. 
The findings are generalized for oscillators with a Cauchy-Lorentzian distribution of natural frequencies. 
\end{abstract}

\maketitle
\thispagestyle{firstpage}

A description of complex systems in terms of a few relevant variables
(order parameters) is an indispensable tool in the theoretical analysis
of equilibrium and nonequilibrium dynamics. In many cases, such a reduction
is possible close to a bifurcation point, where a separation of timescales can be employed to derive closed equations for a few order parameters. In this context,
a seminal breakthrough has been achieved by Ott and Antonsen (OA)~\cite{ott_antonsen_2008}
 for populations of coupled oscillators. They found an exact low-dimensional
reduction of the underlying kinetic equations in terms of the leading Kuramoto
order parameter; this reduction is valid globally and is not restricted to a vicinity
of a synchronization transition. One refers to the configuration described by OA as
the OA manifold. 
Since its discovery, the OA {\it Ansatz} has been adopted in
numerous studies of complex systems out of equilibrium, such as
Josephson 
junctions~\cite{Marvel-Strogatz-09,Marvel-Mirollo-Strogatz-09,Vlasov-Pikovsky-13},
theta neurons and QIF neurons~\cite{Luke-Barreto-So-13,Laing-14,montbrio_pazo_roxin_2015,
laing_2015,bick_goodfellow_laing_martens_2020,Goldobin_etal-21}, 
optomechanical arrays~\cite{Pelka_etal-20}, 
Kuramoto-Battogtokh 
chimeras~\cite{Kuramoto-Battogtokh-02,Laing-09,Bordyugov-Pikovsky-Rosenblum-10,
Panaggio-Abrams-15,Omelchenko-18}, etc.

This Letter extends the OA approach by constructing a hierarchy of global exact
finite-dimensional reductions for the same setup. It is suitable for all the above-mentioned applications of OA. This hierarchy includes the OA manifold
as the zero-order case. Although our variables
are not the order parameters of the ensemble, the latter can be easily calculated.
Taking an array of Josephson junctions as an example, we demonstrate that in our approach
the chaotic collective dynamics, which are impossible within the OA 
manifold, are straightforwardly recovered.   

We consider an ensemble of identical phase oscillators $\varphi_n(t)$, 
with common complex-valued forcing $h(t)$ and common real-valued frequency $\omega(t)$:
\begin{equation}
\dot{\varphi}_n = \omega(t) + \text{Im}\big[2 h(t) e^{-\ii \varphi_n}\big]\;.
\label{eq:phase_system}
\end{equation}
In the thermodynamic limit, where the number of 
oscillators goes to infinity, it is natural to 
describe the system with the phase density distribution 
$P(\varphi,t)$, which obeys the kinetic equation
\begin{equation}
\frac{\partial}{\partial t}P+\frac{\partial}{\partial\varphi} \Big((\omega-\ii 
he^{-\ii \varphi}+\ii h^*e^{\ii \varphi})P \Big) = 0\;.
\label{eq:master}
\end{equation}
It is convenient to express the density in terms of Fourier modes $Z_n=\langle e^{\ii n \varphi}\rangle$, which are the circular moments of the distribution: 
$P(\varphi,t)= \frac{1}{2\pi} (-1+ \sum_{n=0}^\infty Z_n (t)e^{-\ii n\varphi}+\text{c.c.})$.
These moments have a clear physical meaning -- they are the Kuramoto-Daido order parameters~\cite{kuramoto_model,daido_1996}, 
characterizing order (synchronization level) in the population. 
A global coupling in the population
is described by these order parameters (mean fields) via a dependence of the driving terms $h,\omega$ on $Z_n$.
For example, in the seminal Kuramoto-Sakaguchi model~\cite{Sakaguchi-Kuramoto-86}, frequency $\omega$ 
is constant and $h\propto Z_1$;
for coupled Josephson junctions, parameter $h$ is fixed but $\omega$ depends on $Z_1$ \cite{Marvel-Strogatz-09,Marvel-Mirollo-Strogatz-09,Vlasov-Pikovsky-13}.
 
The equations for $Z_n$ (with $Z_0 \equiv 1$ by definition), resulting from the kinetic  equation \eqref{eq:master} read
\begin{equation}
\frac{1}{n}\dot{Z}_n = \ii\omega Z_n + h Z_{n-1}-h^*Z_{n+1}\;, \quad n \geq 1\;.
\label{eq:moment_eqs}
\end{equation}

For the analysis below we introduce the moment exponential generating function (EGF) $F(k,t)$, defined by the series:
\begin{equation}
F(k,t) = \sum_{n=0}^\infty Z_n(t) \frac{k^n}{n!}\;, \quad F(0,t)\equiv 1\;.
\label{eq:egf}
\end{equation}
The infinite system \eqref{eq:moment_eqs} can then be recast 
as a partial differential equation (PDE) for $F(k,t)$
\begin{equation}
\frac{\partial}{\partial t}F = \ii \omega k \frac{\partial}{\partial k}F + h k F - h^* k \frac{\partial^2}{\partial k^2}F\;.
\label{eq:gfpde}
\end{equation}

Below we derive a hierarchy of finite-dimensional representations of the dynamics.
We will first represent all the approaches on the basis of the EGF \eqref{eq:egf}, exploring possible
solutions of Eq.~\eqref{eq:gfpde}. After that, we will show what these representations
mean in terms of the moments $Z_n$. 

The seminal OA {\it Ansatz}~\cite{ott_antonsen_2008}
corresponds to an exponential EGF 
\begin{equation}
F(k,t) = e^{kQ(t)}\;,
\label{eq:oagf}
\end{equation}
 this yields the OA equation for the complex variable $Q(t)$:
\begin{equation}
\dot{Q} = \ii \omega Q +h -h^* Q^2\;.
\label{eq:OA_dynamics}
\end{equation}

We now consider two new {\it Ans\"atze} that each generalize the 
exponential solution \eqref{eq:oagf} to an arbitrarily high-dimensional manifold. 

{\bf Ansatz 1: }
Since the PDE~\eqref{eq:gfpde} is linear, 
we can extend the OA {\it Ansatz}~\eqref{eq:oagf} 
to an arbitrarily large sum of exponentials:
\begin{equation}
F(k,t)=\sum_{m=1}^M \beta_{0,m} e^{k Q_m(t)}\;,
\label{eq:ans1}
\end{equation}
where $\beta_{0,m}$ are complex constants, satisfying the normalization condition 
\begin{equation}
\sum_{m=1}^M \beta_{0,m} =1\;. 
\label{eq:normans1}
\end{equation}
A substitution of Eq.~\eqref{eq:ans1} in \eqref{eq:gfpde} yields a system of $M$ identical equations of form \eqref{eq:OA_dynamics} for the complex variables $Q_m(t)$.
This is the first family of finite-dimensional invariant manifolds. 

{\bf Ansatz 2: }
First, we represent the EGF \eqref{eq:egf} as a product
of an exponential \eqref{eq:oagf} and a leftover function which we denote by $G(k,t)$:
\begin{equation}
F(k,t) = e^{kQ(t)} G(k,t)\;,\quad G(0,t)\equiv 1\;.
\label{eq:ansatz2}
\end{equation}
Next, we suppose that the variable $Q$ satisfies the OA equation \eqref{eq:OA_dynamics}. 
Substituting Eqs.~\eqref{eq:ansatz2},\eqref{eq:OA_dynamics} in Eq.~\eqref{eq:gfpde} then reveals a PDE for $G(k,t)$:
\begin{equation}
\frac{\partial}{\partial t}G = \big[ \ii \omega k - 2h^*Qk \big] \frac{\partial}{\partial k}G - h^* k \frac{\partial^2}{\partial k^2}G\;.
\label{eq:Gpde}
\end{equation}
Equation \eqref{eq:Gpde} has trivial solutions $G = 1$ and $G = e^{k \beta}$, which do
not provide new solutions of the original problem. 
To find nontrivial solutions, let us represent the function $G$ as an infinite series
$G(k,t) = \sum_{r=0}^\infty \beta_{r}(t) \frac{k^r}{r!}$,
where $\beta_{r}$ are complex variables ($\beta_0\equiv 1$). 
Substituting this into Eq. \eqref{eq:Gpde} begets an infinite set of equations for $\beta_r$:
\begin{equation}
\frac{1}{r}\dot{\beta}_{r} = \ii\omega \beta_{r} -2h^*Q\beta_{r}-h^*\beta_{r+1}\;, \quad r \geq 1\;.
\label{eq:betaeqs}
\end{equation}
Because an equation for $\beta_r$ does not contain $\beta_{r-1}$ on the r.h.s. (only terms 
$\sim \beta_r,\beta_{r+1}$ are present), any truncation $\beta_{r>R} = 0 \quad \forall R \in \mathbb{N}$ is invariant. 
The interesting finite-dimensional solutions are therefore those where $G(k,t)$ is a polynomial and
the EGF has the following form:
\begin{equation}
F(k,t) = e^{kQ(t)} \sum\limits_{r=0}^{R} \beta_{r}(t) \frac{k^r}{r!} \;.
\label{eq:egfans2}
\end{equation}
The $Q$ variable evolves according to Eq.~\eqref{eq:OA_dynamics} and the $\beta_r$ variables according to $R$ equations \eqref{eq:betaeqs}, 
where $\beta_{R+1} = 0$. 
This is the second family of finite-dimensional invariant manifolds. 

{\bf Combining Ans\"atze 1 and 2: }
The two {\it Ans\"atze} \eqref{eq:ans1} and \eqref{eq:egfans2} can be combined due to the linearity
of Eq.~\eqref{eq:gfpde}, yielding
\begin{equation}
F(k,t) = \sum\limits_{m=1}^M e^{kQ_m(t)} \sum\limits_{r=0}^{R_m} \beta_{r,m}(t) \frac{k^r}{r!}\;,
\label{eq:ansatze_combined}
\end{equation}
with normalization \eqref{eq:normans1}.
The finite-dimensional set of equations for $\sum_{m=1}^M (1+R_m)$ complex variables $Q_m,\beta_{r,m}$ reads:
\begin{subequations}
\begin{align}
\dot{Q}_m &= \ii\omega Q_m +h -h^*Q_m^2 \;,\label{eq:eqq}\\
\frac{1}{r}\dot{\beta}_{r,m} &= \ii\omega \beta_{r,m} -2h^*Q_m\beta_{r,m}-h^*\beta_{r+1,m}\;.
\label{eq:eqb}
\end{align}\label{eq:the_equations}
\end{subequations}
This combines both approaches and constitutes the extended family of finite-dimensional reductions, which is the main result of this Letter.   

We now establish relations between the new variables $Q_m,\beta_{r,m}$ and the moments $Z_n$. 
For the OA case, Eqs.~\eqref{eq:oagf} and \eqref{eq:egf} yield $Z_n=Q^n$.
Similarly, for {\it Ansatz} 1~\eqref{eq:ans1}, 
$Z_n=\sum_{m=1}^M \beta_{0,m} Q_m^n$.
Less trivial is the expression for moments corresponding to the {\it Ansatz} 2~\eqref{eq:egfans2}. A factorization of an EGF
with an exponential term \eqref{eq:ansatz2} has already been explored in mathematical literature~\cite{number_theory_1993,number_theory_1994}, where 
this factorization has been shown to correspond to a so-called modified binomial transform
\begin{equation}
Z_n = \sum\limits_{r=0}^R \binom{n}{r} \beta_{r} Q^{n-r}\;.
\label{eq:bintr}
 \end{equation}
Combining both expressions,  we obtain a general representation
for the moments in terms of new variables:
\begin{equation}
Z_n = \sum\limits_{m=1}^M \sum\limits_{r=0}^{R_m} \binom{n}{r} \beta_{r,m} Q_m^{n-r}\;. 
\label{eq:moments_combined}
\end{equation}
We stress here, that the derivation of the hierarchy above is not based on any restrictions or assumptions.
Indeed, one can prove by direct calculations \cite{SM} that expressions \eqref{eq:moments_combined} and \eqref{eq:the_equations} solve the original system \eqref{eq:moment_eqs}.

In the Supplemental Material \cite{SM} we additionally relate our approach to circular cumulants, which have been recently suggested for describing the vicinity of the OA manifold~\cite{tyulkina_goldobin_klimenko_pikovsky_2018,tyulkina_goldobin_klimenko_pikovsky_2019}. In contradistinction to the variables $Q_m,\beta_{r,m}$ above,  the hierarchy of cumulants cannot be truncated beyond the first one~\cite{goldobin_dolmatova_2019}. 

It is instructive to look at the phase distributions corresponding to different finite-dimensional 
invariant subspaces. Using moments~\eqref{eq:moments_combined}  to calculate the density, 
we can express $P(\varphi)$ as a sum of elemental real-valued 
contributions corresponding to individual variables $\beta_{r,m}$:
\begin{equation}
P(\varphi) = -\frac{1}{2\pi} + \sum\limits_{m=1}^M \sum\limits_{r=0}^{R_m} P_{r,m}(\varphi)\;,
\label{eq:beta_density}
\end{equation}
where
\begin{equation}
P_{r,m}(\varphi) = \frac{1}{2\pi} \Big(\frac{\beta_{r,m}e^{-\ii r \varphi}}{(1-Q_m e^{-\ii \varphi})^{r+1}} + \text{c.c.} \Big)\,.
\label{eq:elemental_contributions}
\end{equation}
Observe that for $Q_m = 0$ these simply reduce to Fourier modes. Because of the constant $-\frac{1}{2\pi}$ term in Eq.~\eqref{eq:beta_density},
modes $P_{r,m}(\varphi)$ can be interpreted as densities only up to a constant offset, even if the mean value is nonzero.

Only terms with $r=0$ contribute to normalization~\eqref{eq:normans1}. 
They correspond to constants $\beta_{0,m}$ and have a mean value of $\frac{\text{Re}[\beta_{0,m}]}{\pi}$, which can be positive or negative. All other contributions have zero mean. 
The case with $r=0$ and real-valued $\beta_{0,m} \in \mathbb{R}$, corresponds to an offset wrapped Cauchy-Lorentz distribution (WCLD). 
A single WCLD represents the OA formulation, and recently superpositions of several WCLDs were also considered in Refs.~\cite{tyulkina_goldobin_klimenko_pikovsky_2019,engelbrecht_mirollo_2020}. 
The case with $r=0$ and complex $\beta_{0,m} \in \mathbb{C}$, is an offset Kato-Jones distribution (KJD)~\cite{kato-jones}, which is an asymmetric generalization of the WCLD.
Notice that a single complex $\beta_{0,m}$ is not permitted by condition~\eqref{eq:normans1}, which means that a single KJD is not invariant under evolution 
(for a single moment in time it can be represented with two modes: $\beta_{0,1} + \beta_{0,2} = 1$, $Q_1 = 0$). Remarkably, a superposition of KJDs has recently been derived for a seemingly unrelated case of multiharmonic coupling~\cite{Toenjes-Pikovsky-20}. 
Modes \eqref{eq:elemental_contributions} with $r>0$ typically have $r$ local maxima, and to the best of our knowledge, 
have not been studied before. 
In Fig. 1 we depict some examples of modes \eqref{eq:elemental_contributions}; for more see Supplemental Material~\cite{SM} where we also express them in terms of real quantities only (absolute values and arguments). 

The overall distribution \eqref{eq:beta_density} has to be non-negative and normalized, thus imposing restrictions on $Q_m, \beta_{r,m}$ variables. 
While normalization is ensured by condition~\eqref{eq:normans1}, non-negativity presents conditions that are not easily expressed in general (see special cases in Supplemental Material~\cite{SM}, including the case of a single KJD~\cite{kato-jones}).

\begin{figure}[!htbp]
\centering
\includegraphics[width=1.0\columnwidth]{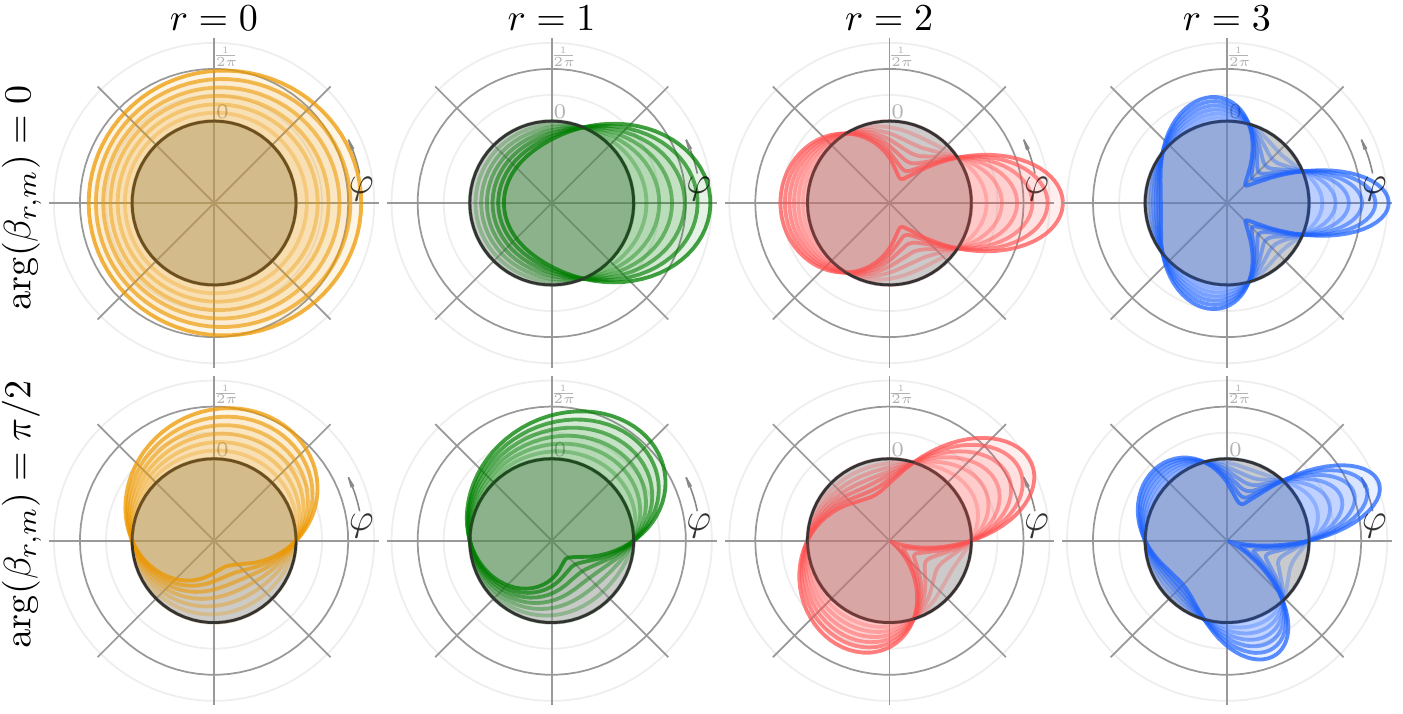}
\caption{
Examples of elemental density contributions $P_{r,m}(\varphi)$~\eqref{eq:elemental_contributions}, for $Q_m = 0.2$, depicted in polar coordinates. 
The first row corresponds to real-valued $\beta_{r,m}$ and its first entry is an offset WCLD which [for full density~\eqref{eq:beta_density}] corresponds to the OA manifold. 
The second row corresponds to purely imaginary $\beta_{r,m}$ and its first entry is an offset KJD with mean zero since $\beta_{r,m}$ has no real component. All other entries have zero mean by definition. 
In each subplot, different curves represent different absolute values of the corresponding variable $|\beta_{r,m}|$. Bold black circle represents zero and the gray disc inside holds negative values (origin is -0.25). Thin outer circle marks the $\frac{1}{2\pi}$ value, representing a uniform normalized distribution for reference. }
\label{fig:dist}
\end{figure} 

On the found finite-dimensional manifolds, one can start with proper
initial conditions in terms of variables $Q_m,\beta_{r,m}$ and follow the finite-dimensional
dynamics according to Eqs.~\eqref{eq:the_equations}. 
The inverse problem -- how to determine
whether a given initial distribution (e.g., given by its order parameters $Z_n$)
lies on a low-dimensional manifold -- is more difficult,
because one needs to invert the infinite systems of Eqs.~\eqref{eq:moments_combined}. 
In the case of just {\it Ansatz} 2, binomial transform \eqref{eq:bintr} can be inverted:
\begin{equation}
\beta_r = \sum_{n=0}^r \binom{r}{n} Z_n (-Q)^{r-n}\;,
\label{eq:ibt}
\end{equation}
but the choice of $Q$ is generally not obvious. 
However, for an initial distribution  with a finite number of nonzero order parameters: $Z_{n>N}(t=0) = 0$
we can set $Q(t=0)=0$ in Eq.~\eqref{eq:ibt} and obtain a finite number of modes  $\beta_r{\footnotesize (t=0)} = Z_r{\footnotesize (t=0)}$. Thus, the evolution of an $N$-moment distribution can be described with $N$ equations of system \eqref{eq:betaeqs} plus one equation for $Q$ \eqref{eq:OA_dynamics}. 
However, finite-$N$ distributions as a class are not invariant under evolution, since the number 
of nonvanishing moments $Z_n$ becomes infinite immediately when $Q\neq0$. 

Because Fourier modes form a base, we conclude that for any initial distribution one can 
find an approximative finite-dimensional description based on the Fourier approximation of the initial condition. 
One might formulate the problem of efficiently approximating a given distribution 
with a finite number of general modes \eqref{eq:elemental_contributions}; however, this problem 
lies beyond the scope of this Letter. 

Up to now we considered a population of identical oscillators \eqref{eq:phase_system}.
In typical situations however, one or several parameters of the oscillators and of the coupling
are distributed. Most popular are models with a distribution of natural frequencies $\omega$,
although other cases have also been considered in the 
literature~\cite{Montbrio-Pazo-11a,Pazo-Montbrio-11,Vlasov-Macau-Pikovsky-14,Iatsenko_etal-14}. 
For definiteness let us consider a distribution of frequencies $g(\omega)$. All the expressions above are valid
for a distribution $P(\varphi,t\,|\,\omega)$, conditioned on a particular value 
of $\omega$: one has $Z_n(t\,|\,\omega)$, 
$Q_m(t\,|\,\omega)$, $\beta_{r,m}(t\,|\,\omega)$, etc. 
Global order parameters are obtained by averaging the latter over $\omega$. 
For example, in the Kuramoto model with coupling via the first global order parameter, the latter
is defined as $\mathcal{Z}_1(t)=\int Z_1(t\,|\,\omega) g(\omega) d\omega$. This expression, 
together with a set of equations~\eqref{eq:the_equations} conditioned under set values of $\omega$, 
constitute a set of integro-differential equations describing the problem.

A particular simplification of this system is possible for a Cauchy-Lorentzian distribution
of frequencies $g(\omega)=\frac{1}{\pi}\frac{\gamma}{(\omega-\omega_0)^2+\gamma^2}$. In this case, 
one can integrate over $g(\omega)$ assuming analyticity of variables $Q_m,\beta_{r,m}$ in the upper 
complex $\omega$ half-plane,
and the integration (using residue theorem) results in a value at the pole $\omega=\omega_0+\ii\gamma$. 
Here the argumentation is essentially the same as used by OA \cite{ott_antonsen_2008}.
As a result, one obtains an exact finite-dimensional description in terms of global variables
$\mathcal{Q}_m(t)=Q_m(t\,|\,\omega_0+\ii\gamma)$, $\mathcal{B}_{r,m}(t)=\beta_{r,m}(t\,|\,\omega_0+\ii\gamma)$ 
(which via~\eqref{eq:moments_combined} relate to global moments $\mathcal{Z}_n$):
\begin{equation}
\begin{aligned}
\dot{\mathcal{Q}}_m&=(\ii \omega_0-\gamma) \mathcal{Q}_m\;
+h-h^* \mathcal{Q}_m^2\;,\\
\frac{1}{r}\dot{\mathcal{B}}_{r,m}&=(\ii\omega_0-\gamma) \mathcal{B}_{r,m}-2 h^*\mathcal{Q}_m
\mathcal{B}_{r,m}-h^*\mathcal{B}_{r+1,m}\;.
\end{aligned}
\label{eq:opomg}
\end{equation}
One can see that in finite representations $\mathcal{B}_{r>R_m,m} = 0$, the last term $\mathcal{B}_{R_m,m}$ does not 
contain driving, so one expects 
that it eventually vanishes because of dissipation $\sim \gamma$. Then, there is no driving
for the next to last $\mathcal{B}_{R_m-1,m}$, and this term will vanish as well, and so on for all $\mathcal{B}_{r>0,m}$ ($\mathcal{B}_{0,m}$ are constants). 
What is left are the $\mathcal{Q}_m$ variables, which are driven by the same force $h$, and will therefore with dissipation eventually converge 
$\lim_{t\to\infty}(\mathcal{Q}_m-\mathcal{Q}_n) = 0,\ \; \forall\; m,n$. 
Thus, one expects 
the OA manifold $\mathcal{B}_{r>0,m}=0,\; \mathcal{Q}_m = \mathcal{Q},\; \mathcal{Z}_n = \mathcal{Q}^n$, to be attractive in this situation, as has been already argued in Refs.~\cite{Ott-Antonsen-09,engelbrecht_mirollo_2020,pietras_daffertshofer_2016}.
As a final note we mention that equations \eqref{eq:opomg} are valid not only for a Cauchy-Lorentzian
distribution of frequencies, but also for oscillators driven by Cauchy 
noise~\cite{Tanaka-20,Toenjes-Pikovsky-20}, where noise strength takes the role of $\gamma$.

\begin{figure}
\centering
\includegraphics[width=0.72\columnwidth]{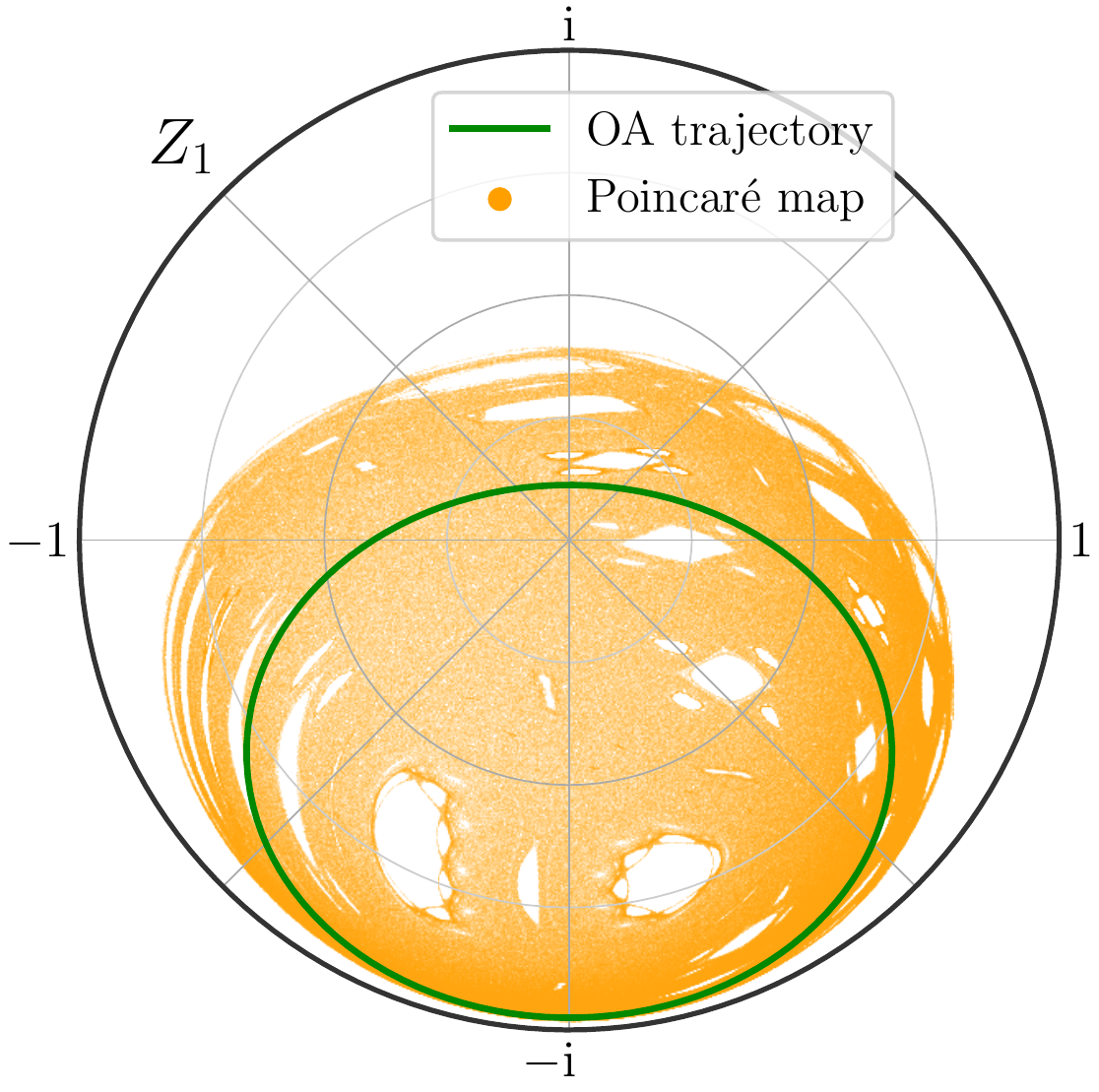}
\caption{Poincar\'e map (orange dots) for dynamics \eqref{eq:jj2} of the Josephson junctions array for $a=1.5$, $\epsilon=-0.7$ and initial condition $Z_1{\footnotesize (t=0)} = 0.4$, $Z_{n>1}{\footnotesize (t=0)} = 0$. The section is made according
to the condition $\arg(\beta)=\pi$, $\frac{d}{dt}\arg(\beta)>0$. Green line: the 
trajectory on the OA manifold with the same initial $Z_1$. }
\label{fig:jj}
\end{figure}

Now we perform a simple numerical simulation to showcase the dynamics on an extended invariant manifold and compare them to the dynamics on the OA manifold. We consider an array of overdamped Josephson junctions coupled via
a resistive load~\cite{Watanabe-Strogatz-94}. It is described by the equations for the Josephson phases
\begin{equation}
\dot\varphi_n=1 + a\sin(\varphi_n)+\frac{\epsilon}{N}\sum_{m=1}^N\sin(\varphi_m)\;.
\label{eq:jj1}
\end{equation}
The OA description of Eq.~\eqref{eq:jj1} yields only regular solutions (because the dimension of the subspace 
is 2~\cite{Martens_etal-09}). Below we show that already within the lowest-order finite-dimensional reduction involving 2 complex variables, chaotic regimes are possible. 
System \eqref{eq:jj1} belongs to the class \eqref{eq:phase_system} with 
$\omega=1+\epsilon\,\text{Im}\langle Z_1 \rangle$ and $h=-\frac{a}{2}$. 
On the two-variable manifold of {\it Ansatz} 2~\eqref{eq:egfans2}, the  equations  in terms of $Q,\beta_1\equiv\beta$ are
\begin{equation}
\begin{aligned}
 \dot Q\ =&\ \ii Q(1+\epsilon\, \text{Im}[Q+\beta])-\frac{a}{2}(1-Q^2)\;,\\
 \dot\beta\ =&\ \ii\beta (1+\epsilon\, \text{Im}[Q+\beta])+a\,Q\beta\;.
\end{aligned}
 \label{eq:jj2}
\end{equation} 
In Fig. \ref{fig:jj} we demonstrate chaotic behavior 
in this system for $a=1.5$, $\epsilon=-0.7$ [Runge-Kutta 4th-order method with time step $2 \times 10^{-3}$ was used;
see Supplemental Material~\cite{SM} for a corresponding simulation
of a large finite ensemble \eqref{eq:jj1}]. Of the four Lyapunov exponents, two are nonzero 
with values $\approx \pm 0.01$. 
For $\beta=0$ (i.e., on the OA manifold) all the solutions are periodic.

There are other examples where the extended finite-dimensional dynamics go beyond the ones
possible in the OA formulation. In Supplemental Material~\cite{SM} we present simulations for a two-population 
chimera~\cite{abrams_mirollo_strogatz_wiley_2008}. There, on the OA manifold (which  is known for such populations of identical oscillators to be only neutrally 
stable~\cite{engelbrecht_mirollo_2020}) the dynamics
of the asynchronous population are periodic, while already the second-order dynamics yield quasiperiodicity
(cf. Refs.~\cite{pikovsky_rosenblum_2008,tyulkina_goldobin_klimenko_pikovsky_2018,tyulkina_goldobin_klimenko_pikovsky_2019}).

Finally, we shortly discuss whether the exact invariant manifolds survive the addition of independent
Gaussian white noise terms to the dynamics of the phases (as mentioned above, for 
Cauchy white noise the invariant manifolds remain valid). In this case, additional terms appear
in equations for moments~\eqref{eq:moment_eqs} and in the PDE for the EGF \eqref{eq:gfpde}.  
An attempt of {\it Ansatz} 2~\eqref{eq:ansatz2} results in an infinite system 
for $\beta_r$ which, in contradistinction to system~\eqref{eq:betaeqs}, cannot be truncated, see Supplemental Material~\cite{SM} for details. Nevertheless, 
truncation might yield an approximative
finite-dimensional description for ensembles with noise; this is a subject
of a forthcoming research (for an approximation in terms of circular cumulants see Ref.~\cite{tyulkina_goldobin_klimenko_pikovsky_2019}).

Summarizing, we have generalized the finite-dimensional description for populations of 
coupled oscillators due to Ott and Antonsen~\cite{ott_antonsen_2008} by constructing 
a family of exact finite-dimensional invariant manifolds for the dynamics. Our approach is as general
and as restricted as that of OA \cite{ott_antonsen_2008}: it is applicable to 
phase dynamics with harmonic forcing and in the thermodynamic limit only, 
but can be generalized to ensembles with distributed frequencies (or other parameters), to oscillators driven by Cauchy noise, etc. It is fully applicable
to systems previously analyzed in terms of the OA reduction, such as Josephson 
junctions~\cite{Marvel-Strogatz-09,Marvel-Mirollo-Strogatz-09,Vlasov-Pikovsky-13},
theta neurons and QIF neurons~\cite{Luke-Barreto-So-13,Laing-14,montbrio_pazo_roxin_2015,
laing_2015,bick_goodfellow_laing_martens_2020,Goldobin_etal-21}, and to Kuramoto-Battogtokh 
chimera~\cite{Kuramoto-Battogtokh-02,Laing-09,Bordyugov-Pikovsky-Rosenblum-10,
Panaggio-Abrams-15,Omelchenko-18}. In situations where the OA manifold is stable,
the extended reductions describe relaxation to this manifold in an exact manner. This description is applicable if the population
is driven away from the OA manifold, e.g., by a phase resetting. In situations where the OA manifold is only neutrally
stable, the extended reductions provide a family of exact equations beyond the OA {\it Ansatz} (there is no condition of a small vicinity, like in the cumulant description \cite{tyulkina_goldobin_klimenko_pikovsky_2018}). Although the analysis of general initial distributions and determining
whether or not they lie on one of the finite-dimensional manifolds remains a challenging problem, we have found
an important and simple class of initial states that do allow for an exact finite-dimensional description: these are
the states with a finite number of nonvanishing moments (Kuramoto-Daido order parameters).

There exists an alternative approach to a finite-dimensional description of oscillator populations,
the Watanabe-Strogatz theory (WS)~\cite{watanabe_strogatz_1993}. While its relation to the OA theory
has been clarified in the literature~\cite{pikovsky_rosenblum_2008,Marvel-Mirollo-Strogatz-09}, a
connection to the present approach remains a subject for future studies. As the WS theory generally
predicts partial integrability for populations of identical oscillators, we can conclude that any high-dimensional dynamics within the found invariant manifolds have a large number of integrals of motion. 

\acknowledgments
The authors thank Denis Goldobin, Lev Smirnov, Ralf Toenjes, Michael Rosenblum, Oleh Omel'chenko, and Erik Mau for useful discussions. The work was supported by DFG (Grant No. PI 220/21-1).


%

\end{document}